\documentclass[aps,reprint,twocolumn,superscriptaddress,10pt]{revtex4-2}

\usepackage[dvipdfmx]{graphicx}
\usepackage{siunitx}
\usepackage{amsfonts}
\usepackage{amsmath}
\usepackage{amssymb}
\usepackage{bm}
\usepackage{graphicx}
\usepackage{dcolumn}
\usepackage{mciteplus}
\usepackage{euscript}
\usepackage{multirow}
\usepackage{textcomp}
\usepackage{gensymb}
\usepackage{float}
\usepackage{enumitem}
\usepackage{xcolor}
\usepackage{sepfootnotes}
\usepackage{mhchem}
\usepackage{outlines}
\usepackage[hidelinks]{hyperref}
\usepackage{lipsum}

\graphicspath{{./}{../../ke2025_light-R_anisotropy/figs/data_for_plot/}{../figs/}}

\DeclareSIUnit{\atomicunit}{a.u.}

\newcommand{\req}[1]{Eq.~\eqref{#1}}

\newcommand{\rfig}[1]{Fig.~\ref{#1}}
\newcommand{\rFig}[1]{Figure~\ref{#1}}

\newcommand{\etp}{E(\theta,\phi)}

\newcommand{\avg}[1]{\langle{#1}\rangle}

\newcommand{\st}[1]{\lvert{#1}\rangle}

\newcommand{\tbvsn}{\text{TbV}_6\text{Sn}_6}
\newcommand{\tbco}{\text{TbCo}_5}
\newcommand{\rvsn}{R\text{V}_6\text{Sn}_6}

\newcommand{\rco}{R\text{Co}_5}

\newcommand{\ar}[2]{A_{#1}^{#2}\avg{r^{#1}}}

\newcommand{\qor}[1]{\frac{Q_{#1}}{\avg{r^{#1}}}}

\begin{document}

\title{Accurate calculation of light rare-earth magnetic anisotropy with density functional theory }

\author{Liqin Ke}
\email{liqin.ke@virginia.edu}
\affiliation{Ames National Laboratory, U.S.~Department of Energy, Ames, IA 50011}
\affiliation{Department of Materials Science and Engineering, University of Virginia, Charlottesville, VA 22904}
\author{R. Flint}
\affiliation{Ames National Laboratory, U.S.~Department of Energy, Ames, IA 50011}
\affiliation{Department of Physics and Astronomy, Iowa State University, Ames, IA 50011}
\author{Y. Lee}
\affiliation{Ames National Laboratory, U.S.~Department of Energy, Ames, IA 50011}

\date{\today} 

\begin{abstract}
Density functional theory (DFT) has long struggled to treat light rare-earth magnetism.
We show that this difficulty arises from an overestimate of the $4f$ charge asphericity, and thus the magnetic anisotropy energy, due to the inadequacy of single Slater-determinant representations.
We propose an effective solution by combining constrained DFT+U with crystal field theory and a systematic many-body correction to the charge asphericity.
We confirm the validity of this combination on TbV$_6$Sn$_6$ and TbCo$_5$, and then show how the many-body correction adjusts the calculated magnetic anisotropy energy of SmCo$_5$
to match experiment.
Our method is an efficient DFT-based approach to address light-rare-earth magnetism.
\end{abstract}


\maketitle

The highly localized $4f$ electrons of the rare earths (RE) are essential not only for fundamental science---from topological magnets~\cite{yin2020n,ma2021prl} to heavy fermions~\cite{Pagliuso2002,Chen2017,Kirchner2020}---but also for technological applications, from high-performance permanent ferromagnets~\cite{skomski1999book,gschneidner1978book,schuler1979pb,lewis2013mmta,mccallum2014armr} to quantum transduction and memory~\cite{Tittel2025}.
These materials have large spin-orbit coupling (SOC) compared to the small crystal field (CF) scales, which allows for uniquely strong magnetocrystalline anisotropy (MA).
The light REs ($n_{4f} < 7$, $R$ = Ce--Sm) are particularly desirable in combination with $3d$ transition metals (TM), as the antiferromagnetic $3d$-$4f$ exchange takes place through the spin, allowing the total moment, $J = L - S$ to align with the $3d$ moments, forming the strongest and most anisotropic permanent magnets, notably SmCo$_5$ and Nd$_2$Fe$_{14}$B.

Despite the broad success of density functional theory (DFT) in many material systems, the open-shell $4f$ materials remain a persistent challenge~\cite{lee2025ncm,zhou2009prb,patrick2018prl} due to the importance of strong interactions.
Interactions may be better accounted for with other methods, but these are computationally expensive and struggle to capture the relatively small CF energy scales~\cite{locht2016,delange2017,lee2025ncm}.
This challenge hinders the systematic, theory-driven design and discovery of RE-based materials.
We have recently addressed one of the fundamental problems: the failure of DFT-based methods to reproduce the correct $4f$ Hund's rule ground state~\cite{zhou2009prb,lee2025ncm}, where we showed that using constrained DFT to enforce the correct Hund's rule ground state allows DFT+U to capture the correct MA across a wide range of \emph{heavy} pristine and doped RE-TM magnets~\cite{lee2025ncm,riberolles2022prx,lee2023prb,rosenberg2022prb,xu2024jacs,han2024a,gazzah2025a}.
Light RE materials provide an additional difficulty, as the SOC $J = L - S$ ground state cannot be represented by a single Slater determinant, while the DFT ground and metastable states will always be single Slater determinants.

In this paper, we address this long-overlooked problem and show that DFT's single Slater determinant nature systematically overestimates the $4f$ charge asphericity, leading to large errors in orbital-dependent quantities like MA.
Fortunately, we show that this overestimation can be systematically countered with a many-body correction that depends only on the RE ion, not on specific material details, and argue that with this correction, DFT can, in fact, be successfully combined with CF theory to accurately treat magnetism throughout the REs.
We propose a systematic method to probe the CF parameters (CFP) using DFT, and validate it in two distinct compounds before applying it to SmCo$_5$, where it allows DFT to reproduce the experimentally determined MA energy.
We expect this approach to work for all integer-valent RE materials, and the obtained CFPs can be widely used, both for low-temperature properties and in the excited spin-orbit multiplets important for optical transitions in both heavy and light REs.

\begin{figure*}[hbtp]
  \includegraphics[width=0.95\linewidth,clip,angle=0]{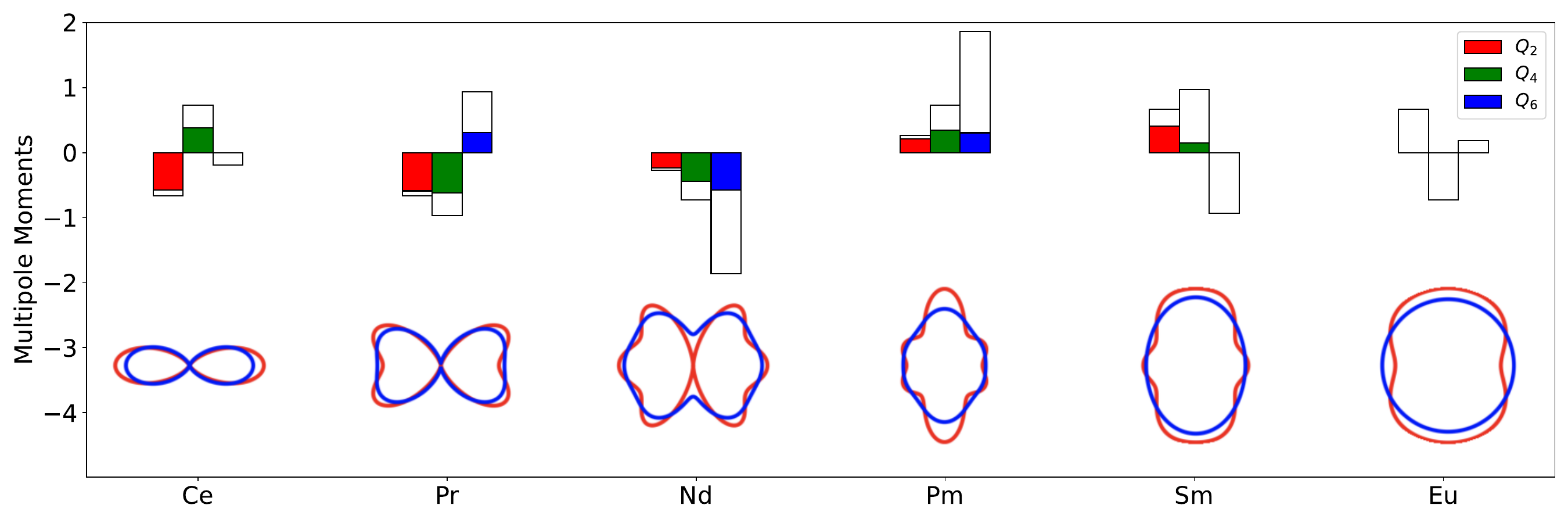}
  \caption{
DFT overestimates the $4f$ charge asphericity of the $\st{J = L - S, M_J = J}$ states for light-$R^{3+}$ ions.  
Polar plots show the true (blue) and DFT (red) $\theta$-dependent charge density $\rho(\theta)$.  
The multipole moments $\tilde{Q}_k = \qor{k}$ ($k = 2$, 4, 6) are shown for the true (solid bars) and DFT (empty bars) charge densities; these are projections of  
$\rho(\theta)$ onto the (unnormalized) Legendre polynomials $P_k^0(\cos\theta)$.  
The overestimation generally increases with both atomic number and multipole order $k$.
  }
\label{fig:q_dq}
\end{figure*}

Multi-$f$-electron RE ions can be described in the Russell-Saunders (LS) coupling scheme, where the total $L$ and $S$ of a given 4$f^n$ configuration are first maximized and then aligned by SOC to give the total $J = L \pm S$ Hund's rule ground state, where $\pm$ corresponds to $n_{4f} \gtrless 7$.
The LS coupling scheme is expected to work well for the REs, while intermediate or $j$-$j$ coupling schemes will be more appropriate for the actinides.
In the LS scheme, the $\st{JM_J}$ states are written as the linear combination,
\begin{equation}
  \st{JM_J} = \!\!\!\!\sum_{M_L + M_S = M_J} \!\!\!\! \langle L S; M_L M_S\vert J M_J\rangle \st{LS; M_L M_S},
  \label{eq:SD}
\end{equation}
where $\langle L S; M_L M_S\vert J M_J\rangle$ are the Clebsch-Gordan coefficients.
The many-body basis states, $\st{LS; M_LM_S}$ are themselves superpositions of single particle states with $M_{L/S}= \sum_{i=1}^{n_{4f}} m_{i,\ell/s}$, but will be represented by a single Slater determinant for the maximally polarized $\st{LS; LS}$ state.
$\st{JM_J}$ can only be represented by a single Slater determinant for $\st{JJ}$, and only when $J = L+S$.

Current DFT implementations can, at best, select only the largest term of this superposition, as they generally represent valence $4f$ electrons by spherical harmonics $\langle \hat r\st{m_\ell} = Y_\ell^m(\theta,\phi)$.
Here, we compare the angular $4f$ charge densities obtained from a single (DFT) versus multiple (true) Slater determinant approximation for all the light $R^{3+}$ ions; this anisotropy can be quantified by the three multipole moments, $Q_k = \int d^3 \rho(\vec{r}) P_k^0(\vec{r})$ ($k=2,4,6$), which are \emph{always} overestimated by DFT.
Here, $P_k^q(\vec{r})$ are Legendre polynomials that proportional to the spherical harmonics, but are not normalized~\cite{abragam1970}.
The MA is directly influenced by the $4f$ charge distribution, which changes its electrostatic interactions with the surrounding material as it is rotated.
Therefore, an incorrect $4f$ charge distribution translates into an incorrect MA energy.
The consequences of this information loss manifest clearly in the angular part of the $4f$ charge density, $\rho(\theta)$ of the fully polarized state, as shown in \rFig{fig:q_dq}.
For Ce$^{3+}$, this approximation is not too bad, as $\st{\frac{5}{2}\, \frac{5}{2} } = \sqrt{\frac{6}{7}}\st{3\, \frac{1}{2}; 3\, \frac{-1}{2}} - \sqrt{\frac{1}{7}}\st{3\, \frac{1}{2}; 2\, \frac{1}{2}}$ and only $1/7$ of the state is dropped.
However, for Eu$^{3+}$, $\st{3\, 3;0\, 0} = \sqrt{\frac{1}{7}} \sum_{M_L = -3}^3 (-1)^L \st{3\, 3; M_L\, -M_L}$, it misses out on $6/7$ of the state and mistakenly predicts anisotropic magnetic behavior as it effectively admixes non-zero $J$ multiplets into the $J = 0$ ground state.

Thus, DFT on its own cannot produce a fully accurate $4f$ state for the light-REs --- even with orbital constraints, the $4f$ wave-function is distorted, and without the Hund's rule constraints, it is entirely wrong for any open-shell $R^{3+}$ ion. 
Nevertheless, we argue that the computed anisotropy energy $E(\theta,\phi)$ from DFT can still be used to extract the true magnetic anisotropy by combining DFT with CF theory.

In CF theory, the spherically symmetric ground state determined by interactions and SOC experiences an anisotropic local environment captured by the
effective Hamiltonian,
\begin{equation}
H_\text{CF} = \sum_{kq} B_k^q O_k^q,
\end{equation}
where $O_k^q$ are Stevens operators written in terms of $\vec{J}$, and $B_k^q$ are a set of often phenomenologically defined CF parameters that depend both
on the ion and its environment.
It is convenient to rewrite $B_k^q = A_k^q \langle r^k\rangle_{4f} \theta_k$, where $\langle r^k\rangle_{4f} \theta_k$ depends not only on the ion but
also on the multiplet given by $(L, S, J)$.
The Stevens coefficients $\theta_k$ are commonly written as $\alpha_J$, $\beta_J$, and $\gamma_J$ for $k = 2, 4, 6$, respectively, and are tabulated for all the
REs~\cite{bethe1929adp,stevens1952pss,ballhausen1962book,hutchings1964ssp,taylor1972book,sievers1982,skomski1999book,skomski2008book}.

CF theory assumes that the $A_k^q$ parameters are independent of both the ion and the $4f$ configuration, and that a single set of parameters can capture
the experimentally observed behavior over a wide range of temperatures and magnetic fields, during which the $4f$ occupancies change substantially; the CF
parameters $A_k^q$ are even thought to remain roughly constant throughout isostructural series of RE compounds.
If the crystal potential is indeed independent of the $4f$ configuration, we should be able to probe it with whichever non-ground-state $4f$ charge
distribution we have.
While both CF theory and DFT are widely used, it has never before been tested whether the anisotropy of DFT calculations can truly be captured by CF
theory, or where this assumption might break down.

For concreteness, we discuss the case of hexagonal symmetry, with $C_{3v}$ point-group symmetry at the RE site, as relevant for $R$Co$_5$. Here, there are generically four parameters: $A_2^0, A_4^0, A_6^0, A_6^6$.  It is useful to define the MA energy as the angle dependent energy of the fully polarized ferromagnetic ground state,  $E(\theta,\phi) = \;_{\hat n}\langle JJ|H_\text{CF} |JJ\rangle_{\hat n}$, where  $|JJ\rangle_{\hat n}$ is along the quantization axis, $\hat n$, rotated by $\theta, \phi$ from the crystallographic $c$-axis.
\begin{align}
E(\theta,\phi) = & \kappa_2^0 P_2(\cos \theta) + \kappa_4^0 P_4(\cos \theta)  + \kappa_6^0 P_6(\cos \theta) \cr
& + \kappa_6^6 \frac{1}{16} \sin^6 \theta \cos 6\phi,
\label{eq:E_kappa}
\end{align}
where $\kappa_\ell^m = A_\ell^m Q_\ell$, $Q_\ell$ depends on $\st{J, M_J}$, and the coefficients and angular dependence are determined using the Wigner
$D$-matrices.
All terms depend on the polar angle and contribute to the uniaxial MA, while only $\kappa_6^6$ has $\phi$-dependence (six-fold in-plane MA)~\cite{han2024a}.

Now we can introduce our DFT-based method to accurately calculated the CFP and MA throughout the REs.
The process consists of three steps, detailed below:
\begin{itemize}
\item Probe the crystal potential within DFT+U by calculating the total energy $E(\theta,\phi)$.
\item Check that CF theory is self-consistent within DFT for the system.
\item Obtain CFPs and apply a many-body correction to the DFT-derived anisotropies.
\end{itemize}

We calculate the total energy $E(\theta,\phi)$ using constrained DFT+$U$, ensuring that the $4f$ occupancy remains in the designated Hund's rule configuration as the spin is rotated.
We then perform a least-squares fit of the calculated $\etp$ to \req{eq:E_kappa} to extract the anisotropy parameters, $\kappa_\ell^m$ and subsequently determine the CFPs, $\ar{\ell}{m}$.
Note that for DFT+$U$ calculations, sizable $U$ values in the range of \SIrange{6}{12}{\eV} are applied to the $R$-$4f$ shell to polarize the occupied and unoccupied $4f$ states and remove them from the Fermi level.
These large $U$'s ensure that the $4f$ states are well-localized and integer valent, with minimal hybridization with the conduction electrons.
Additional details of the DFT+$U$ calculations, including the $U$ dependence, can be found in the Supplemental Material and in Ref.~[~\onlinecite{lee2025ncm}].

For the heavy REs, $\st{JJ}$ is a single Slater determinant and the CFPs calculated within constrained DFT can be used directly.  For light REs, we assume (and justify below) that the DFT ground state probes the same CFPs , but the obtained $A_k^q$'s must be corrected for the overestimation of the $4f$ charge asphericity.  This many-body correction is done using the ratio of the three charge multipole moments, $Q_k$ in the  true many-body ground state, $\st{JJ}$, and the DFT states.  In a given $4f$ configuration, $|GS\rangle$ the multipoles can be written as  $Q^{GS}_k = \langle r^k \rangle_{4f}\langle GS|\theta_k O_k^0|GS\rangle$.
The true many-body ground state is $\st{JJ}$, while only the $\st{LL}$ part of the DFT state is required; $\theta_k$ must be the appropriate Stevens coefficient, $\theta_J$ or $\theta_L$, respectively~\cite{abragam1970,hoffmann1991jpa,duros2025jpa}.
The ratio of the two is 
\begin{equation}
\frac{Q_k}{Q_k^{DFT}} = \frac{\theta_{J} \langle JJ |O_k^0|JJ\rangle}{\theta_{L} \langle LL|O_k^0|LL\rangle}.
\end{equation}
We have tabulated these ratios for the ground state multiplets of all light $R^{3+}$ ions in Table \ref{tbl:ratio}; these can also be calculated for excited multiplets if needed.

\begin{table}[hbt]
  \caption{
    Correction ratios of the true, many-body $Q_k$ to DFT single-Slater determinant $Q^{DFT}_k$ multipole moments for light $R^{3+}$ ions.
  }
\bgroup \def\arraystretch{1.3}
\begin{tabular*}{\linewidth}{c c c@{\extracolsep{\fill}} r r r }
\hline\hline
$R^{3+}$ & J, L, S&  $Q_2/Q_2^{DFT}$ & $Q_4/Q_4^{DFT}$ & $Q_6/Q_6^{DFT}$ \\
\hline
Ce & $\frac{5}{2}, 3, \frac{1}{2}$ & $6/7$ & $11/21$ & $0$ \\
Pr  & $4, 5, 1$ & $728/825$ & $7/11$ & $272/825$ \\
Nd & $\frac{9}{2}, 6, \frac{3}{2}$ & $105/121$ & $952/1573$ & $969/3146$ \\
Pm  & $4, 6, 2$ & $98/121$ & $6664/14157$ & $1292/7865$ \\
Sm & $\frac{5}{2}, 5, \frac{5}{2}$ & $13/21$ & $13/84$ & $0$ \\
Eu  & $0, 3, 3$ & $0$ & $0$ & $0$ \\
\hline
\end{tabular*}
\egroup
\label{tbl:ratio}
\end{table}

Our procedure relies on the consistency of CF theory within DFT: the idea that the DFT single Slater determinant configuration and the true many-body configuration are experiencing the same CF potential, which is not guaranteed.
This validity can be checked by extracting and comparing the CFP for different meta-stable DFT configurations.
While the strength of this consistency may depend on the material, we do expect it to be similar throughout series of RE compounds.
Below, we check this for two well-characterized hexagonal compounds.

\begin{figure}[bhtp]
  \includegraphics[width=.99\linewidth,clip]{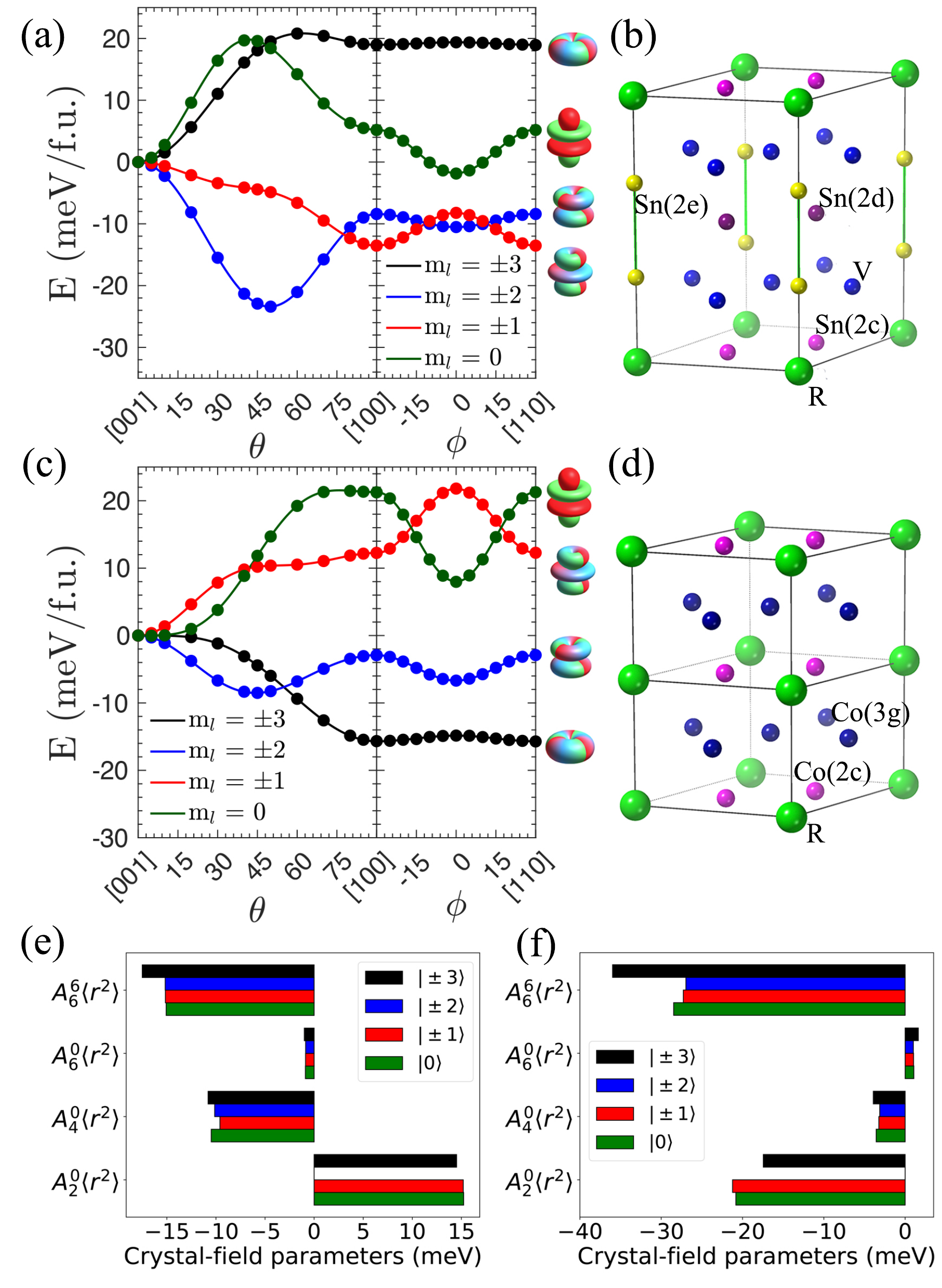}
\caption{
Rare-earth magnetocrystalline anisotropy and crystal-field parameters calculated using various Tb-$4f$ configurations in $\tbvsn$ and $\tbco$.
(a) Magnetic energy variation, $\etp$, in $\tbvsn$ as a function of spin-axis rotation calculated using DFT (dots) for all 
$\st{m_\ell}$ configurations of the single minority-spin electron ($4f^{1\downarrow}$) in Tb$^{3+}$. Insets show polar plots of the corresponding  
$4f^{1\downarrow}$ wavefunctions, with phase denoted by color.  
Solid lines represent CF model fits using \req{eq:E_kappa}.  
The true ground state $\st{m_\ell = 3}$ yields the correct giant easy-axis anisotropy.  
(b) Crystal structure of $\rvsn$.  
(c, d) Same as (a, b), but for $\tbco$.  
(e, f) Crystal-field parameters (CFPs) extracted from the $\etp$ profiles of the $\st{m_\ell}$ states shown in (a, c), demonstrating consistency across different  
$\st{m_\ell}$ probes.  
  }  
  \label{fig:7ml}
\end{figure}

$\tbvsn$ has a crystal structure closely related to that of $\rco$ magnets and exhibits giant easy-axis anisotropy at low
temperatures~\cite{rosenberg2022prb,pokharel2022prm,han2024a}.
It provides a clean test case because the transition-metal sublattice (the V kagome layers) is non-magnetic, so all anisotropy originates from the
Tb$^{3+}$ ions.
In DFT, Tb$^{3+}$ ($4f^8$) is represented as seven $4f$ electrons in the majority spin channel and one in the minority spin channel
($4f^{7\uparrow}4f^{1\downarrow}$).
We calculate $E(\theta,\phi)$ for each choice of the single minority-spin $4f$ electron’s $\st{m_\ell}$, with $m_\ell = -3, \ldots, +3$, thereby probing the
CFPs using effectively four distinct $4f$ charge distributions.

As shown in \rfig{fig:7ml}(a), the DFT+$U$ results for $\etp$ at various $\st{m_\ell}$ values are perfectly fit by the CF model using \req{eq:E_kappa}.
Since $\st{\pm m_\ell}$ states have identical $\rho(\theta)$, we average their calculated $E(\theta,\phi)$ curves (see Supplemental Material).
Note that the Hund’s-rule ground state, $\st{m_\ell = 3}$ yields the expected giant easy-axis MA, consistent with experiment, while other $\st{m_\ell}$
configurations produce drastically different MA curves.

While each $\etp$ profile can be used to extract the CFPs, the response of different $\st{m_\ell}$ states is dominated by different CFPs due to their
distinct $Q^{(m_\ell)}_k = \langle r^k \rangle_{4f}\langle \ell m_\ell|\theta_k O_k^0| \ell m_\ell\rangle$.
Most notably, $Q_2^{(2)} = 0$, meaning that the MA for $|m_\ell = \pm 2\rangle$ originates solely from higher-order ($k = 4,6$) terms.
Indeed, due to orbital-dependent self-interaction errors in DFT, this vanishing $Q_2^{(2)}$ erroneously favors $\st{4f^{1\downarrow}_{m_\ell=2}}$ as the Tb$^{3+}$ ground
state~\cite{lee2023prb}.
Similarly, as $Q^{(3)}_6$ is relatively small, the weak in-plane MA in $\st{m_\ell = \pm 3}$ likely leads to errors in estimating $A_6^6$, which can be
better calculated from $\st{m_\ell = 0}$.

Now, we turn to $R$Co$_5$, where the Co atoms are also magnetic.
As seen in \rfig{fig:7ml}(c), TbCo$_5$ exhibits a much smaller fourth-order term $A_4^0$ than TbV$_6$Sn$_6$, with the uniaxial MA now dominated by the
lowest-order term, $A_2^0$.
$A_2^0$ has the opposite sign compared to that in $\tbvsn$, explaining the easy-axis MA of $\tbvsn$\cite{rosenberg2022prb,pokharel2022prm,riberolles2022prx}
and the easy-plane MA of TbCo$_5$~\cite{greedan1973jssc,zhao1991prb}.
This sign difference likely originates from structural differences: as shown in \rfig{fig:7ml}(b) and (d), in $\rvsn$, a pair of ``dumbbell'' Sn ($2e$) atoms effectively replace an $R$ site in $\rco$ along the $z$-axis and are located closer to the neighboring $R$.
Overall, as we show later, the calculated CFPs can reproduce the experimental anisotropy~\cite{larson2003prb} very well.

Crucially, despite the very different $E(\theta,\phi)$ curves for each $m_\ell$, the extracted CFPs are very similar across all cases for a given material.
For $\tbvsn$, $A_k^q$ values extracted from $m_\ell = 0, \pm1, \pm2$ are nearly identical, while those extracted from $\st{m_\ell = \pm 3}$ show slightly larger deviation (within $\sim 14\%$ across all $m_\ell$ cases). 
We note that the deviations in $\tbco$ are more pronounced than those in $\tbvsn$, which may be attributed to stronger hybridization between $4f$ and $3d$ wavefunctions.

This consistency provides a nontrivial confirmation of the validity of CFT within DFT: it demonstrates that the fitted CFPs are not very sensitive to the specific $4f$ charge distribution used to probe them, with the caveat that $m_\ell = \pm 2$ cannot be used to probe $A_2^0$, as $Q_2 = 0$.
Even though crystal-field effects in real materials arise from both electrostatic and hybridization contributions, the excellent agreement between our DFT results and the CF model confirms the validity of extracting CFPs via self-consistent DFT total-energy calculations.

\begin{figure}[bth]
\centering
\begin{tabular}{c}
\includegraphics[width=.85\linewidth,clip]{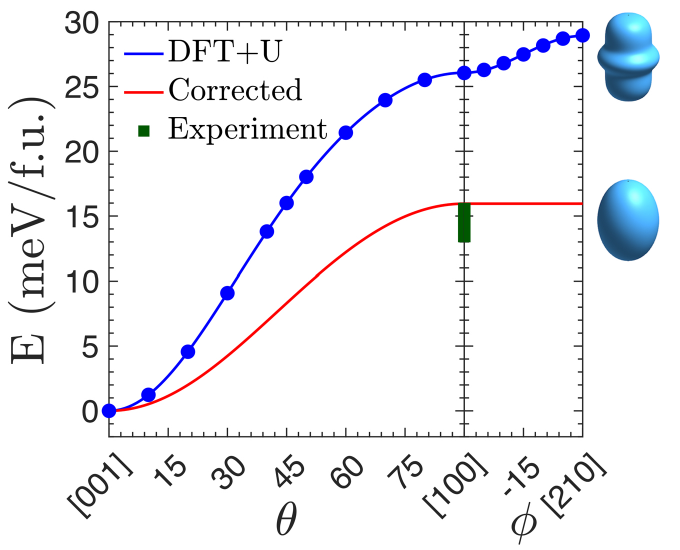}%
\end{tabular}%
\caption{
SmCo$_5$ magnetic anisotropy energy, $\etp$ calculated using DFT+$U$ both without (blue dots/line) and with (red line) many-body corrections.
SmCo$_5$ experimental MAE is in the range of \SIrange{13}{16}{\meV}/f.u. (green bar)~\cite{larson2003prb}. 
The insets show the DFT and corrected $4f$ charge distributions.
}
\label{fig:smco5}
\end{figure}

Now we use these corrections to calculate the MA energy (MAE) of SmCo$_5$.
\rFig{fig:smco5} compares the calculated $\etp$ for SmCo$_5$ using DFT+$U$ ($U = 10$ eV), with and without the many-body correction.
Uncorrected DFT+$U$ predicts a uniaxial MAE of $\sim25$ meV per formula unit, almost double the experimental value of 13--16 meV/f.u.~\cite{larson2003prb}.
After applying our many-body correction, the MAE is significantly reduced, falling within the experimental range.
The correction is also essential for understanding the lack of in-plane MA: DFT predicts a sizable in-plane MA, which vanishes completely after applying the many-body correction, as $Q_6 = 0$ for $J = 5/2$.
Experimentally, this anisotropy can be restored if there is mixing with the excited $J = 7/2$ multiplet.
As another example, in NdCo$_5$, DFT is expected to overestimate the in-plane MA by roughly a factor of $Q_6^{\rm DFT} / Q_6 \approx 3$.

We expect this approach to be broadly applicable to calculating RE crystal fields in both insulating and metallic and both magnetically ordered and paramagnetic
materials; however, there are some limitations.
Integer valence is required, as otherwise the effect of strong hybridization means DFT is no longer probing the material CFP.
It is possible that DFT probes ``bare'' CFPs that can be renormalized by hybridization and interactions, but this requires further investigation.
It is not clear how well this approach will work even for integer-valent Ce materials, where there can be substantial $4f$-$4f$ hopping if the Ce-Ce
distance is less than the Hill limit.
While we have only tested this on ground-state $J$ multiplets, the CFPs should not be substantially different for excited $J$ multiplets with the same
total $L$ and $S$, which may allow for the calculation of optical transitions.
Actinide materials will provide additional challenges, both due to the more extended $5f$ wavefunctions leading to stronger hybridization and $f$-$f$
hopping, and the different SO coupling schemes required, but it may well be possible to extend this method to some insulating actinide materials.

\emph{In summary}, we have identified a fundamental issue in single-Slater-determinant DFT calculations for light REs: an exaggerated $4f$ charge
asphericity leads to an overestimation of the MAE. To resolve this issue, we presented a combined DFT and CF theory approach that uses constrained DFT+$U$
to probe the CFPs, which we have shown to be largely independent of the $4f$ electronic configuration. These DFT parameters are then corrected by
$Q_k/Q_k^{\rm DFT}$ to obtain the true CFPs and MAE.
With this correction, our DFT+$U$ calculations for SmCo$_5$ now show close agreement with the experimental MAE. This many-body correction is especially
important for heavier light-REs like Sm and Eu, and in cases where higher-order CF terms play a significant role.
Higher-order CF terms are important in materials with substantial in-plane MA; those with large high-order contributions to the uniaxial MAE, as
in $\rvsn$; and in cubic materials, where the first anisotropy arises from $Q_4$. In cubic Sm$^{3+}$ materials, the many-body correction
$Q_4/Q_4^{\rm DFT}$ reduces the $4f$ MA to approximately 15\% of the DFT value.
More broadly, the self-consistent constrained-DFT approach demonstrated here offers a practical and systematic method to extract CFPs and incorporate many-body effects in $4f$ magnetism, paving the way for more accurate and efficient modeling of RE materials.

\vspace{12pt}

\emph{Acknowledgments}---
This work is supported by the U.S.~Department of Energy, Office of Basic Energy Sciences, Materials Sciences and Engineering Division.
The initial conceptual development by Ke was supported by the Office of Science Early Career Research Program through the Office of Basic Energy Sciences, Materials Sciences and Engineering Division.
Ames Laboratory is operated for the U.S.~Department of Energy by Iowa State University under Contract No.~DE-AC02-07CH11358.
We acknowledge Research Computing at the University of Virginia for providing computational resources.

\bibliography{aaa}

\begin{thebibliography}{36}%
\makeatletter
\providecommand \@ifxundefined [1]{%
 \@ifx{#1\undefined}
}%
\providecommand \@ifnum [1]{%
 \ifnum #1\expandafter \@firstoftwo
 \else \expandafter \@secondoftwo
 \fi
}%
\providecommand \@ifx [1]{%
 \ifx #1\expandafter \@firstoftwo
 \else \expandafter \@secondoftwo
 \fi
}%
\providecommand \natexlab [1]{#1}%
\providecommand \enquote  [1]{``#1''}%
\providecommand \bibnamefont  [1]{#1}%
\providecommand \bibfnamefont [1]{#1}%
\providecommand \citenamefont [1]{#1}%
\providecommand \href@noop [0]{\@secondoftwo}%
\providecommand \href [0]{\begingroup \@sanitize@url \@href}%
\providecommand \@href[1]{\@@startlink{#1}\@@href}%
\providecommand \@@href[1]{\endgroup#1\@@endlink}%
\providecommand \@sanitize@url [0]{\catcode `\\12\catcode `\$12\catcode
  `\&12\catcode `\#12\catcode `\^12\catcode `\_12\catcode `\%12\relax}%
\providecommand \@@startlink[1]{}%
\providecommand \@@endlink[0]{}%
\providecommand \url  [0]{\begingroup\@sanitize@url \@url }%
\providecommand \@url [1]{\endgroup\@href {#1}{\urlprefix }}%
\providecommand \urlprefix  [0]{URL }%
\providecommand \Eprint [0]{\href }%
\providecommand \doibase [0]{https://doi.org/}%
\providecommand \selectlanguage [0]{\@gobble}%
\providecommand \bibinfo  [0]{\@secondoftwo}%
\providecommand \bibfield  [0]{\@secondoftwo}%
\providecommand \translation [1]{[#1]}%
\providecommand \BibitemOpen [0]{}%
\providecommand \bibitemStop [0]{}%
\providecommand \bibitemNoStop [0]{.\EOS\space}%
\providecommand \EOS [0]{\spacefactor3000\relax}%
\providecommand \BibitemShut  [1]{\csname bibitem#1\endcsname}%
\let\auto@bib@innerbib\@empty
\bibitem [{\citenamefont {Yin}\ \emph {et~al.}(2020)\citenamefont {Yin},
  \citenamefont {Ma}, \citenamefont {Cochran}, \citenamefont {Xu},
  \citenamefont {Zhang}, \citenamefont {Tien}, \citenamefont {Shumiya},
  \citenamefont {Cheng}, \citenamefont {Jiang}, \citenamefont {Lian} \emph
  {et~al.}}]{yin2020n}%
  \BibitemOpen
  \bibfield  {author} {\bibinfo {author} {\bibfnamefont {J.-X.}\ \bibnamefont
  {Yin}}, \bibinfo {author} {\bibfnamefont {W.}~\bibnamefont {Ma}}, \bibinfo
  {author} {\bibfnamefont {T.~A.}\ \bibnamefont {Cochran}}, \bibinfo {author}
  {\bibfnamefont {X.}~\bibnamefont {Xu}}, \bibinfo {author} {\bibfnamefont
  {S.~S.}\ \bibnamefont {Zhang}}, \bibinfo {author} {\bibfnamefont {H.-J.}\
  \bibnamefont {Tien}}, \bibinfo {author} {\bibfnamefont {N.}~\bibnamefont
  {Shumiya}}, \bibinfo {author} {\bibfnamefont {G.}~\bibnamefont {Cheng}},
  \bibinfo {author} {\bibfnamefont {K.}~\bibnamefont {Jiang}}, \bibinfo
  {author} {\bibfnamefont {B.}~\bibnamefont {Lian}}, \emph {et~al.},\
  }\bibfield  {title} {\bibinfo {title} {{Quantum-limit Chern topological
  magnetism in TbMn$_6$Sn$_6$}},\ }\href@noop {} {\bibfield  {journal}
  {\bibinfo  {journal} {Nature}\ }\textbf {\bibinfo {volume} {583}},\ \bibinfo
  {pages} {533} (\bibinfo {year} {2020})}\BibitemShut {NoStop}%
\bibitem [{\citenamefont {Ma}\ \emph {et~al.}(2021)\citenamefont {Ma},
  \citenamefont {Xu}, \citenamefont {Yin}, \citenamefont {Yang}, \citenamefont
  {Zhou}, \citenamefont {Cheng}, \citenamefont {Huang}, \citenamefont {Qu},
  \citenamefont {Wang}, \citenamefont {Hasan},\ and\ \citenamefont
  {Jia}}]{ma2021prl}%
  \BibitemOpen
  \bibfield  {author} {\bibinfo {author} {\bibfnamefont {W.}~\bibnamefont
  {Ma}}, \bibinfo {author} {\bibfnamefont {X.}~\bibnamefont {Xu}}, \bibinfo
  {author} {\bibfnamefont {J.-X.}\ \bibnamefont {Yin}}, \bibinfo {author}
  {\bibfnamefont {H.}~\bibnamefont {Yang}}, \bibinfo {author} {\bibfnamefont
  {H.}~\bibnamefont {Zhou}}, \bibinfo {author} {\bibfnamefont {Z.-J.}\
  \bibnamefont {Cheng}}, \bibinfo {author} {\bibfnamefont {Y.}~\bibnamefont
  {Huang}}, \bibinfo {author} {\bibfnamefont {Z.}~\bibnamefont {Qu}}, \bibinfo
  {author} {\bibfnamefont {F.}~\bibnamefont {Wang}}, \bibinfo {author}
  {\bibfnamefont {M.~Z.}\ \bibnamefont {Hasan}},\ and\ \bibinfo {author}
  {\bibfnamefont {S.}~\bibnamefont {Jia}},\ }\bibfield  {title} {\bibinfo
  {title} {{Rare Earth Engineering in $R$Mn$_6$Sn$_6$ ($R=$Gd--Tm, Lu)
  Topological Kagome Magnets}},\ }\href
  {https://doi.org/10.1103/PhysRevLett.126.246602} {\bibfield  {journal}
  {\bibinfo  {journal} {Phys. Rev. Lett.}\ }\textbf {\bibinfo {volume} {126}},\
  \bibinfo {pages} {246602} (\bibinfo {year} {2021})}\BibitemShut {NoStop}%
\bibitem [{\citenamefont {Pagliuso}\ \emph {et~al.}(2002)\citenamefont
  {Pagliuso}, \citenamefont {Curro}, \citenamefont {Moreno}, \citenamefont
  {Hundley}, \citenamefont {Thompson}, \citenamefont {Sarrao},\ and\
  \citenamefont {Fisk}}]{Pagliuso2002}%
  \BibitemOpen
  \bibfield  {author} {\bibinfo {author} {\bibfnamefont {P.}~\bibnamefont
  {Pagliuso}}, \bibinfo {author} {\bibfnamefont {N.}~\bibnamefont {Curro}},
  \bibinfo {author} {\bibfnamefont {N.}~\bibnamefont {Moreno}}, \bibinfo
  {author} {\bibfnamefont {M.}~\bibnamefont {Hundley}}, \bibinfo {author}
  {\bibfnamefont {J.}~\bibnamefont {Thompson}}, \bibinfo {author}
  {\bibfnamefont {J.}~\bibnamefont {Sarrao}},\ and\ \bibinfo {author}
  {\bibfnamefont {Z.}~\bibnamefont {Fisk}},\ }\bibfield  {title} {\bibinfo
  {title} {Structurally tuned superconductivity in heavy-fermion {Ce$M$In$_5$
  ($M$=Co, Ir, Rh)}},\ }\href
  {https://doi.org/https://doi.org/10.1016/S0921-4526(02)00751-2} {\bibfield
  {journal} {\bibinfo  {journal} {Physica B: Condensed Matter}\ }\textbf
  {\bibinfo {volume} {320}},\ \bibinfo {pages} {370} (\bibinfo {year}
  {2002})},\ \bibinfo {note} {proceedings of the Fifth Latin American Workshop
  on Magnetism, Magnetic Materials and their Applications}\BibitemShut
  {NoStop}%
\bibitem [{\citenamefont {Chen}\ \emph {et~al.}(2017)\citenamefont {Chen},
  \citenamefont {Xu}, \citenamefont {Niu}, \citenamefont {Jiang}, \citenamefont
  {Peng}, \citenamefont {Xu}, \citenamefont {Wen}, \citenamefont {Ding},
  \citenamefont {Huang}, \citenamefont {Shu}, \citenamefont {Zhang},
  \citenamefont {Lee}, \citenamefont {Strocov}, \citenamefont {Shi},
  \citenamefont {Bisti}, \citenamefont {Schmitt}, \citenamefont {Huang},
  \citenamefont {Dudin}, \citenamefont {Lai}, \citenamefont {Kirchner},
  \citenamefont {Yuan},\ and\ \citenamefont {Feng}}]{Chen2017}%
  \BibitemOpen
  \bibfield  {author} {\bibinfo {author} {\bibfnamefont {Q.~Y.}\ \bibnamefont
  {Chen}}, \bibinfo {author} {\bibfnamefont {D.~F.}\ \bibnamefont {Xu}},
  \bibinfo {author} {\bibfnamefont {X.~H.}\ \bibnamefont {Niu}}, \bibinfo
  {author} {\bibfnamefont {J.}~\bibnamefont {Jiang}}, \bibinfo {author}
  {\bibfnamefont {R.}~\bibnamefont {Peng}}, \bibinfo {author} {\bibfnamefont
  {H.~C.}\ \bibnamefont {Xu}}, \bibinfo {author} {\bibfnamefont {C.~H.~P.}\
  \bibnamefont {Wen}}, \bibinfo {author} {\bibfnamefont {Z.~F.}\ \bibnamefont
  {Ding}}, \bibinfo {author} {\bibfnamefont {K.}~\bibnamefont {Huang}},
  \bibinfo {author} {\bibfnamefont {L.}~\bibnamefont {Shu}}, \bibinfo {author}
  {\bibfnamefont {Y.~J.}\ \bibnamefont {Zhang}}, \bibinfo {author}
  {\bibfnamefont {H.}~\bibnamefont {Lee}}, \bibinfo {author} {\bibfnamefont
  {V.~N.}\ \bibnamefont {Strocov}}, \bibinfo {author} {\bibfnamefont
  {M.}~\bibnamefont {Shi}}, \bibinfo {author} {\bibfnamefont {F.}~\bibnamefont
  {Bisti}}, \bibinfo {author} {\bibfnamefont {T.}~\bibnamefont {Schmitt}},
  \bibinfo {author} {\bibfnamefont {Y.~B.}\ \bibnamefont {Huang}}, \bibinfo
  {author} {\bibfnamefont {P.}~\bibnamefont {Dudin}}, \bibinfo {author}
  {\bibfnamefont {X.~C.}\ \bibnamefont {Lai}}, \bibinfo {author} {\bibfnamefont
  {S.}~\bibnamefont {Kirchner}}, \bibinfo {author} {\bibfnamefont {H.~Q.}\
  \bibnamefont {Yuan}},\ and\ \bibinfo {author} {\bibfnamefont {D.~L.}\
  \bibnamefont {Feng}},\ }\bibfield  {title} {\bibinfo {title} {Direct
  observation of how the heavy-fermion state develops in {CeCoIn$_{5}$}},\
  }\href {https://doi.org/10.1103/PhysRevB.96.045107} {\bibfield  {journal}
  {\bibinfo  {journal} {Phys. Rev. B}\ }\textbf {\bibinfo {volume} {96}},\
  \bibinfo {pages} {045107} (\bibinfo {year} {2017})}\BibitemShut {NoStop}%
\bibitem [{\citenamefont {Kirchner}\ \emph {et~al.}(2020)\citenamefont
  {Kirchner}, \citenamefont {Paschen}, \citenamefont {Chen}, \citenamefont
  {Wirth}, \citenamefont {Feng}, \citenamefont {Thompson},\ and\ \citenamefont
  {Si}}]{Kirchner2020}%
  \BibitemOpen
  \bibfield  {author} {\bibinfo {author} {\bibfnamefont {S.}~\bibnamefont
  {Kirchner}}, \bibinfo {author} {\bibfnamefont {S.}~\bibnamefont {Paschen}},
  \bibinfo {author} {\bibfnamefont {Q.}~\bibnamefont {Chen}}, \bibinfo {author}
  {\bibfnamefont {S.}~\bibnamefont {Wirth}}, \bibinfo {author} {\bibfnamefont
  {D.}~\bibnamefont {Feng}}, \bibinfo {author} {\bibfnamefont {J.~D.}\
  \bibnamefont {Thompson}},\ and\ \bibinfo {author} {\bibfnamefont
  {Q.}~\bibnamefont {Si}},\ }\bibfield  {title} {\bibinfo {title} {{Colloquium:
  Heavy-electron quantum criticality and single-particle spectroscopy}},\
  }\href {https://doi.org/10.1103/RevModPhys.92.011002} {\bibfield  {journal}
  {\bibinfo  {journal} {Rev. Mod. Phys.}\ }\textbf {\bibinfo {volume} {92}},\
  \bibinfo {pages} {011002} (\bibinfo {year} {2020})}\BibitemShut {NoStop}%
\bibitem [{\citenamefont {Skomski}\ and\ \citenamefont
  {Coey}(1999)}]{skomski1999book}%
  \BibitemOpen
  \bibfield  {author} {\bibinfo {author} {\bibfnamefont {R.}~\bibnamefont
  {Skomski}}\ and\ \bibinfo {author} {\bibfnamefont {J.~M.~D.}\ \bibnamefont
  {Coey}},\ }\href@noop {} {\emph {\bibinfo {title} {Permanent magnetism}}}\
  (\bibinfo  {publisher} {Institute of Physics Publishing, Bristol},\ \bibinfo
  {year} {1999})\BibitemShut {NoStop}%
\bibitem [{\citenamefont {Gschneidner}\ and\ \citenamefont
  {Eyring}(1978)}]{gschneidner1978book}%
  \BibitemOpen
  \bibinfo {editor} {\bibfnamefont {K.~A.}\ \bibnamefont {Gschneidner}}\ and\
  \bibinfo {editor} {\bibfnamefont {L.}~\bibnamefont {Eyring}},\ eds.,\
  \href@noop {} {\emph {\bibinfo {title} {Handbook on the Physics and Chemistry
  of Rare Earths}}}\ (\bibinfo  {publisher} {North Holland, Amsterdam},\
  \bibinfo {year} {1978})\BibitemShut {NoStop}%
\bibitem [{\citenamefont {Sch\"{u}ler}(1979)}]{schuler1979pb}%
  \BibitemOpen
  \bibfield  {author} {\bibinfo {author} {\bibfnamefont {K.}~\bibnamefont
  {Sch\"{u}ler}},\ }\bibfield  {title} {\bibinfo {title} {Mccaig: Permanent
  magnets in theory and practice},\ }\href@noop {} {\bibfield  {journal}
  {\bibinfo  {journal} {Physikalische Blätter}\ }\textbf {\bibinfo {volume}
  {35}},\ \bibinfo {pages} {423} (\bibinfo {year} {1979})}\BibitemShut
  {NoStop}%
\bibitem [{\citenamefont {Lewis}\ and\ \citenamefont
  {Jiménez-Villacorta}(2013)}]{lewis2013mmta}%
  \BibitemOpen
  \bibfield  {author} {\bibinfo {author} {\bibfnamefont {L.~H.}\ \bibnamefont
  {Lewis}}\ and\ \bibinfo {author} {\bibfnamefont {F.}~\bibnamefont
  {Jiménez-Villacorta}},\ }\bibfield  {title} {\bibinfo {title} {Perspectives
  on permanent magnetic materials for energy conversion and power generation},\
  }\href {https://doi.org/10.1007/s11661-012-1278-2} {\bibfield  {journal}
  {\bibinfo  {journal} {Metallurgical and Materials Transactions A}\ }\textbf
  {\bibinfo {volume} {44}},\ \bibinfo {pages} {20} (\bibinfo {year}
  {2013})}\BibitemShut {NoStop}%
\bibitem [{\citenamefont {McCallum}\ \emph {et~al.}(2014)\citenamefont
  {McCallum}, \citenamefont {Lewis}, \citenamefont {Skomski}, \citenamefont
  {Kramer},\ and\ \citenamefont {{Anderson}}}]{mccallum2014armr}%
  \BibitemOpen
  \bibfield  {author} {\bibinfo {author} {\bibfnamefont {R.}~\bibnamefont
  {McCallum}}, \bibinfo {author} {\bibfnamefont {L.}~\bibnamefont {Lewis}},
  \bibinfo {author} {\bibfnamefont {R.}~\bibnamefont {Skomski}}, \bibinfo
  {author} {\bibfnamefont {M.}~\bibnamefont {Kramer}},\ and\ \bibinfo {author}
  {\bibfnamefont {I.}~\bibnamefont {{Anderson}}},\ }\bibfield  {title}
  {\bibinfo {title} {Practical aspects of modern and future permanent
  magnets},\ }\href {https://doi.org/10.1146/annurev-matsci-070813-113457}
  {\bibfield  {journal} {\bibinfo  {journal} {Annual Review of Materials
  Research}\ }\textbf {\bibinfo {volume} {44}},\ \bibinfo {pages} {451}
  (\bibinfo {year} {2014})}\BibitemShut {NoStop}%
\bibitem [{\citenamefont {Tittel}\ \emph {et~al.}(2025)\citenamefont {Tittel},
  \citenamefont {Afzelius}, \citenamefont {Kinos}, \citenamefont {Rippe},\ and\
  \citenamefont {Walther}}]{Tittel2025}%
  \BibitemOpen
  \bibfield  {author} {\bibinfo {author} {\bibfnamefont {W.}~\bibnamefont
  {Tittel}}, \bibinfo {author} {\bibfnamefont {M.}~\bibnamefont {Afzelius}},
  \bibinfo {author} {\bibfnamefont {A.}~\bibnamefont {Kinos}}, \bibinfo
  {author} {\bibfnamefont {L.}~\bibnamefont {Rippe}},\ and\ \bibinfo {author}
  {\bibfnamefont {A.}~\bibnamefont {Walther}},\ }\bibfield  {title} {\bibinfo
  {title} {Quantum networks using rare-earth ions},\ }\href
  {https://doi.org/10.1088/2058-9565/addd93} {\bibfield  {journal} {\bibinfo
  {journal} {Quantum Science and Technology}\ }\textbf {\bibinfo {volume}
  {10}},\ \bibinfo {pages} {033002} (\bibinfo {year} {2025})}\BibitemShut
  {NoStop}%
\bibitem [{\citenamefont {Lee}\ \emph {et~al.}(2025)\citenamefont {Lee},
  \citenamefont {Ning}, \citenamefont {Flint}, \citenamefont {McQueeney},
  \citenamefont {Mazin},\ and\ \citenamefont {Ke}}]{lee2025ncm}%
  \BibitemOpen
  \bibfield  {author} {\bibinfo {author} {\bibfnamefont {Y.}~\bibnamefont
  {Lee}}, \bibinfo {author} {\bibfnamefont {Z.}~\bibnamefont {Ning}}, \bibinfo
  {author} {\bibfnamefont {R.}~\bibnamefont {Flint}}, \bibinfo {author}
  {\bibfnamefont {R.~J.}\ \bibnamefont {McQueeney}}, \bibinfo {author}
  {\bibfnamefont {I.~I.}\ \bibnamefont {Mazin}},\ and\ \bibinfo {author}
  {\bibfnamefont {L.}~\bibnamefont {Ke}},\ }\bibfield  {title} {\bibinfo
  {title} {Importance of enforcing hund’s rules in density functional theory
  calculations of rare earth magnetocrystalline anisotropy},\ }\href
  {https://doi.org/10.1038/s41524-025-01632-3} {\bibfield  {journal} {\bibinfo
  {journal} {npj Computational Materials}\ }\textbf {\bibinfo {volume} {11}},\
  \bibinfo {pages} {168} (\bibinfo {year} {2025})}\BibitemShut {NoStop}%
\bibitem [{\citenamefont {Zhou}\ and\ \citenamefont
  {Ozoli\c{n}\v{s}}(2009)}]{zhou2009prb}%
  \BibitemOpen
  \bibfield  {author} {\bibinfo {author} {\bibfnamefont {F.}~\bibnamefont
  {Zhou}}\ and\ \bibinfo {author} {\bibfnamefont {V.}~\bibnamefont
  {Ozoli\c{n}\v{s}}},\ }\bibfield  {title} {\bibinfo {title} {Obtaining correct
  orbital ground states in $f$-electron systems using a nonspherical
  self-interaction-corrected {LDA+$U$} method},\ }\href
  {https://doi.org/10.1103/PhysRevB.80.125127} {\bibfield  {journal} {\bibinfo
  {journal} {Phys. Rev. B}\ }\textbf {\bibinfo {volume} {80}},\ \bibinfo
  {pages} {125127} (\bibinfo {year} {2009})}\BibitemShut {NoStop}%
\bibitem [{\citenamefont {Patrick}\ \emph {et~al.}(2018)\citenamefont
  {Patrick}, \citenamefont {Kumar}, \citenamefont {Balakrishnan}, \citenamefont
  {Edwards}, \citenamefont {Lees}, \citenamefont {Petit},\ and\ \citenamefont
  {Staunton}}]{patrick2018prl}%
  \BibitemOpen
  \bibfield  {author} {\bibinfo {author} {\bibfnamefont {C.~E.}\ \bibnamefont
  {Patrick}}, \bibinfo {author} {\bibfnamefont {S.}~\bibnamefont {Kumar}},
  \bibinfo {author} {\bibfnamefont {G.}~\bibnamefont {Balakrishnan}}, \bibinfo
  {author} {\bibfnamefont {R.~S.}\ \bibnamefont {Edwards}}, \bibinfo {author}
  {\bibfnamefont {M.~R.}\ \bibnamefont {Lees}}, \bibinfo {author}
  {\bibfnamefont {L.}~\bibnamefont {Petit}},\ and\ \bibinfo {author}
  {\bibfnamefont {J.~B.}\ \bibnamefont {Staunton}},\ }\bibfield  {title}
  {\bibinfo {title} {Calculating the magnetic anisotropy of
  rare-earth--transition-metal ferrimagnets},\ }\href
  {https://doi.org/10.1103/PhysRevLett.120.097202} {\bibfield  {journal}
  {\bibinfo  {journal} {Phys. Rev. Lett.}\ }\textbf {\bibinfo {volume} {120}},\
  \bibinfo {pages} {097202} (\bibinfo {year} {2018})}\BibitemShut {NoStop}%
\bibitem [{\citenamefont {Locht}\ \emph {et~al.}(2016)\citenamefont {Locht},
  \citenamefont {Kvashnin}, \citenamefont {Rodrigues}, \citenamefont {Pereiro},
  \citenamefont {Bergman}, \citenamefont {Bergqvist}, \citenamefont
  {Lichtenstein}, \citenamefont {Katsnelson}, \citenamefont {Delin},
  \citenamefont {Klautau}, \citenamefont {Johansson}, \citenamefont
  {Di~Marco},\ and\ \citenamefont {Eriksson}}]{locht2016}%
  \BibitemOpen
  \bibfield  {author} {\bibinfo {author} {\bibfnamefont {I.~L.~M.}\
  \bibnamefont {Locht}}, \bibinfo {author} {\bibfnamefont {Y.~O.}\ \bibnamefont
  {Kvashnin}}, \bibinfo {author} {\bibfnamefont {D.~C.~M.}\ \bibnamefont
  {Rodrigues}}, \bibinfo {author} {\bibfnamefont {M.}~\bibnamefont {Pereiro}},
  \bibinfo {author} {\bibfnamefont {A.}~\bibnamefont {Bergman}}, \bibinfo
  {author} {\bibfnamefont {L.}~\bibnamefont {Bergqvist}}, \bibinfo {author}
  {\bibfnamefont {A.~I.}\ \bibnamefont {Lichtenstein}}, \bibinfo {author}
  {\bibfnamefont {M.~I.}\ \bibnamefont {Katsnelson}}, \bibinfo {author}
  {\bibfnamefont {A.}~\bibnamefont {Delin}}, \bibinfo {author} {\bibfnamefont
  {A.~B.}\ \bibnamefont {Klautau}}, \bibinfo {author} {\bibfnamefont
  {B.}~\bibnamefont {Johansson}}, \bibinfo {author} {\bibfnamefont
  {I.}~\bibnamefont {Di~Marco}},\ and\ \bibinfo {author} {\bibfnamefont
  {O.}~\bibnamefont {Eriksson}},\ }\bibfield  {title} {\bibinfo {title}
  {Standard model of the rare earths analyzed from the {Hubbard I}
  approximation},\ }\href {https://doi.org/10.1103/PhysRevB.94.085137}
  {\bibfield  {journal} {\bibinfo  {journal} {Phys. Rev. B}\ }\textbf {\bibinfo
  {volume} {94}},\ \bibinfo {pages} {085137} (\bibinfo {year}
  {2016})}\BibitemShut {NoStop}%
\bibitem [{\citenamefont {Delange}\ \emph {et~al.}(2017)\citenamefont
  {Delange}, \citenamefont {Biermann}, \citenamefont {Miyake},\ and\
  \citenamefont {Pourovskii}}]{delange2017}%
  \BibitemOpen
  \bibfield  {author} {\bibinfo {author} {\bibfnamefont {P.}~\bibnamefont
  {Delange}}, \bibinfo {author} {\bibfnamefont {S.}~\bibnamefont {Biermann}},
  \bibinfo {author} {\bibfnamefont {T.}~\bibnamefont {Miyake}},\ and\ \bibinfo
  {author} {\bibfnamefont {L.}~\bibnamefont {Pourovskii}},\ }\bibfield  {title}
  {\bibinfo {title} {Crystal-field splittings in rare-earth-based hard magnets:
  An ab initio approach},\ }\href {https://doi.org/10.1103/PhysRevB.96.155132}
  {\bibfield  {journal} {\bibinfo  {journal} {Phys. Rev. B}\ }\textbf {\bibinfo
  {volume} {96}},\ \bibinfo {pages} {155132} (\bibinfo {year}
  {2017})}\BibitemShut {NoStop}%
\bibitem [{\citenamefont {Riberolles}\ \emph {et~al.}(2022)\citenamefont
  {Riberolles}, \citenamefont {Slade}, \citenamefont {Abernathy}, \citenamefont
  {Granroth}, \citenamefont {Li}, \citenamefont {Lee}, \citenamefont
  {Canfield}, \citenamefont {Ueland}, \citenamefont {Ke},\ and\ \citenamefont
  {McQueeney}}]{riberolles2022prx}%
  \BibitemOpen
  \bibfield  {author} {\bibinfo {author} {\bibfnamefont {S.~X.~M.}\
  \bibnamefont {Riberolles}}, \bibinfo {author} {\bibfnamefont {T.~J.}\
  \bibnamefont {Slade}}, \bibinfo {author} {\bibfnamefont {D.~L.}\ \bibnamefont
  {Abernathy}}, \bibinfo {author} {\bibfnamefont {G.~E.}\ \bibnamefont
  {Granroth}}, \bibinfo {author} {\bibfnamefont {B.}~\bibnamefont {Li}},
  \bibinfo {author} {\bibfnamefont {Y.}~\bibnamefont {Lee}}, \bibinfo {author}
  {\bibfnamefont {P.~C.}\ \bibnamefont {Canfield}}, \bibinfo {author}
  {\bibfnamefont {B.~G.}\ \bibnamefont {Ueland}}, \bibinfo {author}
  {\bibfnamefont {L.}~\bibnamefont {Ke}},\ and\ \bibinfo {author}
  {\bibfnamefont {R.~J.}\ \bibnamefont {McQueeney}},\ }\bibfield  {title}
  {\bibinfo {title} {{Low-Temperature Competing Magnetic Energy Scales in the
  Topological Ferrimagnet TbMn$_6$Sn$_6$}},\ }\href
  {https://doi.org/10.1103/PhysRevX.12.021043} {\bibfield  {journal} {\bibinfo
  {journal} {Phys. Rev. X}\ }\textbf {\bibinfo {volume} {12}},\ \bibinfo
  {pages} {021043} (\bibinfo {year} {2022})}\BibitemShut {NoStop}%
\bibitem [{\citenamefont {Lee}\ \emph {et~al.}(2023)\citenamefont {Lee},
  \citenamefont {Skomski}, \citenamefont {Wang}, \citenamefont {Orth},
  \citenamefont {Ren}, \citenamefont {Kang}, \citenamefont {Pathak},
  \citenamefont {Kutepov}, \citenamefont {Harmon}, \citenamefont {McQueeney},
  \citenamefont {Mazin},\ and\ \citenamefont {Ke}}]{lee2023prb}%
  \BibitemOpen
  \bibfield  {author} {\bibinfo {author} {\bibfnamefont {Y.}~\bibnamefont
  {Lee}}, \bibinfo {author} {\bibfnamefont {R.}~\bibnamefont {Skomski}},
  \bibinfo {author} {\bibfnamefont {X.}~\bibnamefont {Wang}}, \bibinfo {author}
  {\bibfnamefont {P.~P.}\ \bibnamefont {Orth}}, \bibinfo {author}
  {\bibfnamefont {Y.}~\bibnamefont {Ren}}, \bibinfo {author} {\bibfnamefont
  {B.}~\bibnamefont {Kang}}, \bibinfo {author} {\bibfnamefont {A.~K.}\
  \bibnamefont {Pathak}}, \bibinfo {author} {\bibfnamefont {A.}~\bibnamefont
  {Kutepov}}, \bibinfo {author} {\bibfnamefont {B.~N.}\ \bibnamefont {Harmon}},
  \bibinfo {author} {\bibfnamefont {R.~J.}\ \bibnamefont {McQueeney}}, \bibinfo
  {author} {\bibfnamefont {I.~I.}\ \bibnamefont {Mazin}},\ and\ \bibinfo
  {author} {\bibfnamefont {L.}~\bibnamefont {Ke}},\ }\bibfield  {title}
  {\bibinfo {title} {Interplay between magnetism and band topology in the
  kagome magnets {$R$Mn$_6$Sn$_6$}},\ }\href
  {https://doi.org/10.1103/PhysRevB.108.045132} {\bibfield  {journal} {\bibinfo
   {journal} {Phys. Rev. B}\ }\textbf {\bibinfo {volume} {108}},\ \bibinfo
  {pages} {045132} (\bibinfo {year} {2023})}\BibitemShut {NoStop}%
\bibitem [{\citenamefont {Rosenberg}\ \emph {et~al.}(2022)\citenamefont
  {Rosenberg}, \citenamefont {DeStefano}, \citenamefont {Guo}, \citenamefont
  {Oh}, \citenamefont {Hashimoto}, \citenamefont {Lu}, \citenamefont
  {Birgeneau}, \citenamefont {Lee}, \citenamefont {Ke}, \citenamefont {Yi},\
  and\ \citenamefont {Chu}}]{rosenberg2022prb}%
  \BibitemOpen
  \bibfield  {author} {\bibinfo {author} {\bibfnamefont {E.}~\bibnamefont
  {Rosenberg}}, \bibinfo {author} {\bibfnamefont {J.~M.}\ \bibnamefont
  {DeStefano}}, \bibinfo {author} {\bibfnamefont {Y.}~\bibnamefont {Guo}},
  \bibinfo {author} {\bibfnamefont {J.~S.}\ \bibnamefont {Oh}}, \bibinfo
  {author} {\bibfnamefont {M.}~\bibnamefont {Hashimoto}}, \bibinfo {author}
  {\bibfnamefont {D.}~\bibnamefont {Lu}}, \bibinfo {author} {\bibfnamefont
  {R.~J.}\ \bibnamefont {Birgeneau}}, \bibinfo {author} {\bibfnamefont
  {Y.}~\bibnamefont {Lee}}, \bibinfo {author} {\bibfnamefont {L.}~\bibnamefont
  {Ke}}, \bibinfo {author} {\bibfnamefont {M.}~\bibnamefont {Yi}},\ and\
  \bibinfo {author} {\bibfnamefont {J.-H.}\ \bibnamefont {Chu}},\ }\bibfield
  {title} {\bibinfo {title} {Uniaxial ferromagnetism in the kagome metal
  {TbV$_{6}$Sn$_6$}},\ }\href {https://doi.org/10.1103/PhysRevB.106.115139}
  {\bibfield  {journal} {\bibinfo  {journal} {Phys. Rev. B}\ }\textbf {\bibinfo
  {volume} {106}},\ \bibinfo {pages} {115139} (\bibinfo {year}
  {2022})}\BibitemShut {NoStop}%
\bibitem [{\citenamefont {Xu}\ \emph {et~al.}(2024)\citenamefont {Xu},
  \citenamefont {Lee}, \citenamefont {Ke}, \citenamefont {Kang}, \citenamefont
  {Boswell}, \citenamefont {Bud'ko}, \citenamefont {Zhou}, \citenamefont {Ke},
  \citenamefont {Li}, \citenamefont {Canfield},\ and\ \citenamefont
  {Xie}}]{xu2024jacs}%
  \BibitemOpen
  \bibfield  {author} {\bibinfo {author} {\bibfnamefont {M.}~\bibnamefont
  {Xu}}, \bibinfo {author} {\bibfnamefont {Y.}~\bibnamefont {Lee}}, \bibinfo
  {author} {\bibfnamefont {X.}~\bibnamefont {Ke}}, \bibinfo {author}
  {\bibfnamefont {M.-C.}\ \bibnamefont {Kang}}, \bibinfo {author}
  {\bibfnamefont {M.}~\bibnamefont {Boswell}}, \bibinfo {author} {\bibfnamefont
  {S.~L.}\ \bibnamefont {Bud'ko}}, \bibinfo {author} {\bibfnamefont
  {L.}~\bibnamefont {Zhou}}, \bibinfo {author} {\bibfnamefont {L.}~\bibnamefont
  {Ke}}, \bibinfo {author} {\bibfnamefont {M.}~\bibnamefont {Li}}, \bibinfo
  {author} {\bibfnamefont {P.~C.}\ \bibnamefont {Canfield}},\ and\ \bibinfo
  {author} {\bibfnamefont {W.}~\bibnamefont {Xie}},\ }\bibfield  {title}
  {\bibinfo {title} {{{Giant Uniaxial Magnetocrystalline Anisotropy in
  SmCrGe$_3$}}},\ }\href {https://doi.org/10.1021/jacs.4c10056} {\bibfield
  {journal} {\bibinfo  {journal} {Journal of the American Chemical Society}\
  }\textbf {\bibinfo {volume} {146}},\ \bibinfo {pages} {30294} (\bibinfo
  {year} {2024})}\BibitemShut {NoStop}%
\bibitem [{\citenamefont {Han}\ \emph {et~al.}(2024)\citenamefont {Han},
  \citenamefont {McKenzie}, \citenamefont {Blawat}, \citenamefont {Slade},
  \citenamefont {Lee}, \citenamefont {Pajerowski}, \citenamefont {Singleton},
  \citenamefont {Li}, \citenamefont {Canfield}, \citenamefont {Ke},
  \citenamefont {McDonald}, \citenamefont {Flint},\ and\ \citenamefont
  {McQueeney}}]{han2024a}%
  \BibitemOpen
  \bibfield  {author} {\bibinfo {author} {\bibfnamefont {T.}~\bibnamefont
  {Han}}, \bibinfo {author} {\bibfnamefont {R.~D.}\ \bibnamefont {McKenzie}},
  \bibinfo {author} {\bibfnamefont {J.}~\bibnamefont {Blawat}}, \bibinfo
  {author} {\bibfnamefont {T.~J.}\ \bibnamefont {Slade}}, \bibinfo {author}
  {\bibfnamefont {Y.}~\bibnamefont {Lee}}, \bibinfo {author} {\bibfnamefont
  {D.~M.}\ \bibnamefont {Pajerowski}}, \bibinfo {author} {\bibfnamefont
  {J.}~\bibnamefont {Singleton}}, \bibinfo {author} {\bibfnamefont
  {B.}~\bibnamefont {Li}}, \bibinfo {author} {\bibfnamefont {P.~C.}\
  \bibnamefont {Canfield}}, \bibinfo {author} {\bibfnamefont {L.}~\bibnamefont
  {Ke}}, \bibinfo {author} {\bibfnamefont {R.}~\bibnamefont {McDonald}},
  \bibinfo {author} {\bibfnamefont {R.}~\bibnamefont {Flint}},\ and\ \bibinfo
  {author} {\bibfnamefont {R.~J.}\ \bibnamefont {McQueeney}},\ }\href
  {https://arxiv.org/abs/2412.02010} {\bibinfo {title} {Proximity to quantum
  criticality in the {Ising} ferromagnet {TbV$_6$Sn$_6$}}} (\bibinfo {year}
  {2024}),\ \Eprint {https://arxiv.org/abs/2412.02010} {arXiv:2412.02010
  [cond-mat.str-el]} \BibitemShut {NoStop}%
\bibitem [{\citenamefont {Gazzah}\ \emph {et~al.}(2025)\citenamefont {Gazzah},
  \citenamefont {Chang}, \citenamefont {Lee}, \citenamefont {Bhandari},
  \citenamefont {Regmi}, \citenamefont {Zhou}, \citenamefont {Mitchell},
  \citenamefont {Ke}, \citenamefont {Mazin},\ and\ \citenamefont
  {Ghimire}}]{gazzah2025a}%
  \BibitemOpen
  \bibfield  {author} {\bibinfo {author} {\bibfnamefont {M.~E.}\ \bibnamefont
  {Gazzah}}, \bibinfo {author} {\bibfnamefont {P.-H.}\ \bibnamefont {Chang}},
  \bibinfo {author} {\bibfnamefont {Y.}~\bibnamefont {Lee}}, \bibinfo {author}
  {\bibfnamefont {H.}~\bibnamefont {Bhandari}}, \bibinfo {author}
  {\bibfnamefont {R.}~\bibnamefont {Regmi}}, \bibinfo {author} {\bibfnamefont
  {X.}~\bibnamefont {Zhou}}, \bibinfo {author} {\bibfnamefont {J.~F.}\
  \bibnamefont {Mitchell}}, \bibinfo {author} {\bibfnamefont {L.}~\bibnamefont
  {Ke}}, \bibinfo {author} {\bibfnamefont {I.~I.}\ \bibnamefont {Mazin}},\ and\
  \bibinfo {author} {\bibfnamefont {N.~J.}\ \bibnamefont {Ghimire}},\ }\href
  {https://arxiv.org/abs/2505.02936} {\bibinfo {title} {Doping-induced spin
  reorientation in kagome magnet tmmn6sn6}} (\bibinfo {year} {2025}),\ \Eprint
  {https://arxiv.org/abs/2505.02936} {arXiv:2505.02936 [cond-mat.mtrl-sci]}
  \BibitemShut {NoStop}%
\bibitem [{\citenamefont {Abragam}\ and\ \citenamefont
  {Bleaney}(1970)}]{abragam1970}%
  \BibitemOpen
  \bibfield  {author} {\bibinfo {author} {\bibfnamefont {A.}~\bibnamefont
  {Abragam}}\ and\ \bibinfo {author} {\bibfnamefont {B.}~\bibnamefont
  {Bleaney}},\ }\href@noop {} {\emph {\bibinfo {title} {Electron Paramagnetic
  Resonance of Transition Ions}}},\ \bibinfo {edition} {1st}\ ed.,\
  International Series of Monographs on Physics\ (\bibinfo  {publisher}
  {Clarendon Press},\ \bibinfo {year} {1970})\BibitemShut {NoStop}%
\bibitem [{\citenamefont {Bethe}(1929)}]{bethe1929adp}%
  \BibitemOpen
  \bibfield  {author} {\bibinfo {author} {\bibfnamefont {H.}~\bibnamefont
  {Bethe}},\ }\bibfield  {title} {\bibinfo {title} {Termaufspaltung in
  kristallen},\ }\href@noop {} {\bibfield  {journal} {\bibinfo  {journal}
  {Annalen der Physik}\ }\textbf {\bibinfo {volume} {395}},\ \bibinfo {pages}
  {133} (\bibinfo {year} {1929})}\BibitemShut {NoStop}%
\bibitem [{\citenamefont {Stevens}(1952)}]{stevens1952pss}%
  \BibitemOpen
  \bibfield  {author} {\bibinfo {author} {\bibfnamefont {K.~W.~H.}\
  \bibnamefont {Stevens}},\ }\bibfield  {title} {\bibinfo {title} {Matrix
  elements and operator equivalents connected with the magnetic properties of
  rare earth ions},\ }\href {https://doi.org/10.1088/0370-1298/65/3/308}
  {\bibfield  {journal} {\bibinfo  {journal} {Proceedings of the Physical
  Society. Section A}\ }\textbf {\bibinfo {volume} {65}},\ \bibinfo {pages}
  {209} (\bibinfo {year} {1952})}\BibitemShut {NoStop}%
\bibitem [{\citenamefont {Ballhausen}(1962)}]{ballhausen1962book}%
  \BibitemOpen
  \bibfield  {author} {\bibinfo {author} {\bibfnamefont {C.~J.}\ \bibnamefont
  {Ballhausen}},\ }\href@noop {} {\emph {\bibinfo {title} {Introduction to
  ligand field theory}}}\ (\bibinfo  {publisher} {McGraw-Hill, New York},\
  \bibinfo {year} {1962})\BibitemShut {NoStop}%
\bibitem [{\citenamefont {Hutchings}(1964)}]{hutchings1964ssp}%
  \BibitemOpen
  \bibfield  {author} {\bibinfo {author} {\bibfnamefont {M.}~\bibnamefont
  {Hutchings}},\ }\bibfield  {title} {\bibinfo {title} {{Point-Charge
  Calculations of Energy Levels of Magnetic Ions in Crystalline Electric
  Fields}}\ }(\bibinfo  {publisher} {Academic Press},\ \bibinfo {year} {1964})\
  pp.\ \bibinfo {pages} {227--273}\BibitemShut {NoStop}%
\bibitem [{\citenamefont {Taylor}\ and\ \citenamefont
  {Darby}(1972)}]{taylor1972book}%
  \BibitemOpen
  \bibfield  {author} {\bibinfo {author} {\bibfnamefont {K.~N.~R.}\
  \bibnamefont {Taylor}}\ and\ \bibinfo {author} {\bibfnamefont {M.~I.}\
  \bibnamefont {Darby}},\ }\href@noop {} {\emph {\bibinfo {title} {Physics of
  rare earth solids}}}\ (\bibinfo  {publisher} {Chapman and Hall, London},\
  \bibinfo {year} {1972})\BibitemShut {NoStop}%
\bibitem [{\citenamefont {Sievers}(1982)}]{sievers1982}%
  \BibitemOpen
  \bibfield  {author} {\bibinfo {author} {\bibfnamefont {J.}~\bibnamefont
  {Sievers}},\ }\bibfield  {title} {\bibinfo {title} {{Asphericity of 4f-shells
  in their Hund's rule ground states}},\ }\href
  {https://doi.org/10.1007/BF01321865} {\bibfield  {journal} {\bibinfo
  {journal} {Zeitschrift f{\"u}r Physik B Condensed Matter}\ }\textbf {\bibinfo
  {volume} {45}},\ \bibinfo {pages} {289} (\bibinfo {year} {1982})}\BibitemShut
  {NoStop}%
\bibitem [{\citenamefont {Skomski}(2008)}]{skomski2008book}%
  \BibitemOpen
  \bibfield  {author} {\bibinfo {author} {\bibfnamefont {R.}~\bibnamefont
  {Skomski}},\ }\href@noop {} {\emph {\bibinfo {title} {Simple Models of
  Magnetism}}},\ Oxford Graduate Texts\ (\bibinfo  {publisher} {OUP Oxford},\
  \bibinfo {year} {2008})\BibitemShut {NoStop}%
\bibitem [{\citenamefont {Hoffmann}(1991)}]{hoffmann1991jpa}%
  \BibitemOpen
  \bibfield  {author} {\bibinfo {author} {\bibfnamefont {P.}~\bibnamefont
  {Hoffmann}},\ }\bibfield  {title} {\bibinfo {title} {Generalization of
  stevens' operator-equivalent method},\ }\href
  {https://doi.org/10.1088/0305-4470/24/1/014} {\bibfield  {journal} {\bibinfo
  {journal} {Journal of Physics A: Mathematical and General}\ }\textbf
  {\bibinfo {volume} {24}},\ \bibinfo {pages} {35} (\bibinfo {year}
  {1991})}\BibitemShut {NoStop}%
\bibitem [{\citenamefont {Duros}\ \emph {et~al.}(2024)\citenamefont {Duros},
  \citenamefont {Juhin}, \citenamefont {Elnaggar}, \citenamefont
  {Chiuzbăian},\ and\ \citenamefont {Brouder}}]{duros2025jpa}%
  \BibitemOpen
  \bibfield  {author} {\bibinfo {author} {\bibfnamefont {O.}~\bibnamefont
  {Duros}}, \bibinfo {author} {\bibfnamefont {A.}~\bibnamefont {Juhin}},
  \bibinfo {author} {\bibfnamefont {H.}~\bibnamefont {Elnaggar}}, \bibinfo
  {author} {\bibfnamefont {G.~S.}\ \bibnamefont {Chiuzbăian}},\ and\ \bibinfo
  {author} {\bibfnamefont {C.}~\bibnamefont {Brouder}},\ }\bibfield  {title}
  {\bibinfo {title} {General expressions for stevens and racah operator
  equivalents},\ }\href {https://doi.org/10.1088/1751-8121/ad96fc} {\bibfield
  {journal} {\bibinfo  {journal} {Journal of Physics A: Mathematical and
  Theoretical}\ }\textbf {\bibinfo {volume} {58}},\ \bibinfo {pages} {025207}
  (\bibinfo {year} {2024})}\BibitemShut {NoStop}%
\bibitem [{\citenamefont {Pokharel}\ \emph {et~al.}(2022)\citenamefont
  {Pokharel}, \citenamefont {Ortiz}, \citenamefont {Chamorro}, \citenamefont
  {Sarte}, \citenamefont {Kautzsch}, \citenamefont {Wu}, \citenamefont {Ruff},\
  and\ \citenamefont {Wilson}}]{pokharel2022prm}%
  \BibitemOpen
  \bibfield  {author} {\bibinfo {author} {\bibfnamefont {G.}~\bibnamefont
  {Pokharel}}, \bibinfo {author} {\bibfnamefont {B.}~\bibnamefont {Ortiz}},
  \bibinfo {author} {\bibfnamefont {J.}~\bibnamefont {Chamorro}}, \bibinfo
  {author} {\bibfnamefont {P.}~\bibnamefont {Sarte}}, \bibinfo {author}
  {\bibfnamefont {L.}~\bibnamefont {Kautzsch}}, \bibinfo {author}
  {\bibfnamefont {G.}~\bibnamefont {Wu}}, \bibinfo {author} {\bibfnamefont
  {J.}~\bibnamefont {Ruff}},\ and\ \bibinfo {author} {\bibfnamefont {S.~D.}\
  \bibnamefont {Wilson}},\ }\bibfield  {title} {\bibinfo {title} {Highly
  anisotropic magnetism in the vanadium-based kagome metal {TbV$_6$Sn$_6$}},\
  }\href {https://doi.org/10.1103/PhysRevMaterials.6.104202} {\bibfield
  {journal} {\bibinfo  {journal} {Phys. Rev. Mater.}\ }\textbf {\bibinfo
  {volume} {6}},\ \bibinfo {pages} {104202} (\bibinfo {year}
  {2022})}\BibitemShut {NoStop}%
\bibitem [{\citenamefont {Greedan}\ and\ \citenamefont
  {Rao}(1973)}]{greedan1973jssc}%
  \BibitemOpen
  \bibfield  {author} {\bibinfo {author} {\bibfnamefont {J.}~\bibnamefont
  {Greedan}}\ and\ \bibinfo {author} {\bibfnamefont {V.}~\bibnamefont {Rao}},\
  }\bibfield  {title} {\bibinfo {title} {{An analysis of the rare earth
  contribution to the magnetic anisotropy in $R$Co$_5$ and $R$$_2$Co$_{17}$
  compounds}},\ }\href
  {https://doi.org/https://doi.org/10.1016/0022-4596(73)90228-4} {\bibfield
  {journal} {\bibinfo  {journal} {Journal of Solid State Chemistry}\ }\textbf
  {\bibinfo {volume} {6}},\ \bibinfo {pages} {387} (\bibinfo {year}
  {1973})}\BibitemShut {NoStop}%
\bibitem [{\citenamefont {Tie-song}\ \emph {et~al.}(1991)\citenamefont
  {Tie-song}, \citenamefont {Han-min}, \citenamefont {Guang-hua}, \citenamefont
  {Xiu-feng},\ and\ \citenamefont {Hong}}]{zhao1991prb}%
  \BibitemOpen
  \bibfield  {author} {\bibinfo {author} {\bibfnamefont {Z.}~\bibnamefont
  {Tie-song}}, \bibinfo {author} {\bibfnamefont {J.}~\bibnamefont {Han-min}},
  \bibinfo {author} {\bibfnamefont {G.}~\bibnamefont {Guang-hua}}, \bibinfo
  {author} {\bibfnamefont {H.}~\bibnamefont {Xiu-feng}},\ and\ \bibinfo
  {author} {\bibfnamefont {C.}~\bibnamefont {Hong}},\ }\bibfield  {title}
  {\bibinfo {title} {{Magnetic properties of R ions in $R$Co$_{5}$ compounds
  ($R$=Pr, Nd, Sm, Gd, Tb, Dy, Ho, and Er)}},\ }\href
  {https://doi.org/10.1103/PhysRevB.43.8593} {\bibfield  {journal} {\bibinfo
  {journal} {Phys. Rev. B}\ }\textbf {\bibinfo {volume} {43}},\ \bibinfo
  {pages} {8593} (\bibinfo {year} {1991})}\BibitemShut {NoStop}%
\bibitem [{\citenamefont {Larson}\ \emph {et~al.}(2003)\citenamefont {Larson},
  \citenamefont {Mazin},\ and\ \citenamefont
  {Papaconstantopoulos}}]{larson2003prb}%
  \BibitemOpen
  \bibfield  {author} {\bibinfo {author} {\bibfnamefont {P.}~\bibnamefont
  {Larson}}, \bibinfo {author} {\bibfnamefont {I.~I.}\ \bibnamefont {Mazin}},\
  and\ \bibinfo {author} {\bibfnamefont {D.~A.}\ \bibnamefont
  {Papaconstantopoulos}},\ }\bibfield  {title} {\bibinfo {title} {Calculation
  of magnetic anisotropy energy in {SmCo$_5$}},\ }\href
  {https://doi.org/10.1103/PhysRevB.67.214405} {\bibfield  {journal} {\bibinfo
  {journal} {Phys. Rev. B}\ }\textbf {\bibinfo {volume} {67}},\ \bibinfo
  {pages} {214405} (\bibinfo {year} {2003})}\BibitemShut {NoStop}%
\end{thebibliography}%

\end{document}